\documentclass[11pt,twoside]{article}

\usepackage{graphicx}

\usepackage{asp2006}
\usepackage{lscape}

\begin{document}
\title{Impact of coronal mass ejections, interchange reconnection, and disconnection on heliospheric magnetic field strength}   

\author{N.U. Crooker}   
\affil{Center for Space Physics, Boston University, Boston, MA 02215, USA}    
\author{M.J. Owens}   
\affil{Space and Atmospheric Physics, Imperial College London, UK}    
\affil{Now at: Space Environment Physics Group, Department of Meteorology, University of Reading, UK}

\begin{abstract} 
An update of \citet{owe08} shows that the relationship between the coronal mass ejection (CME) rate and the heliospheric magnetic field strength predicts a field floor of less than 4 nT at 1 AU. This implies that the record low values measured during this solar minimum do not necessarily contradict the idea that open flux is conserved. The results are consistent with the hypothesis that CMEs add flux to the heliosphere and interchange reconnection between open flux and closed CME loops subtracts flux. An existing model embracing this hypothesis, however, overestimates flux during the current minimum, even though the CME rate has been low. The discrepancy calls for reasonable changes in model assumptions.
\end{abstract}

\section{Introduction}
Using an analytical, empirically-based model, \citet{owe06,owe07} demonstrated the feasibility of the idea that the rise and fall of magnetic field strength in the heliosphere over the solar cycle could be caused solely by the addition of closed magnetic loops from coronal mass ejections (CMEs) coupled with a loss of flux through reconnection at the Sun. Either of two patterns of reconnection can satisfy the constraints posed by observations of suprathermal electrons, which can sense magnetic topology: 1) CME loops gradually open by interchange reconnection with open field lines, or 2) open field lines of opposite polarity reconnect, thus disconnecting flux from the Sun.

The two options for flux loss through reconnection have different advantages. Option 1 is attractive because it restricts the rise and fall of field magnitude to a single phenomenon, CMEs, and it conserves open flux, which serves as a backdrop. On the other hand, because option 2 calls upon a phenomenon separate from CMEs for the loss of flux, that is, disconnection elsewhere, it has the flexibility of reducing the heliospheric flux to zero at solar minimum rather than to some backdrop value. In view of the weakness of the heliospheric magnetic field strength during the present solar minimum, unprecedented in the $\sim$50-year history of in situ measurements, we ask whether option 1 is still viable.

A preliminary answer of ``yes'' to the question of option 1 viability was provided by \citet{owe08} using data through 5 July 2008. Since the heliospheric field strength continued to drop after that date, this report updates the results of \citet{owe08} using data through 30 May 2009.

\section{Analysis}
\begin{figure}
\centerline{\includegraphics[width=1.0\textwidth,clip=]{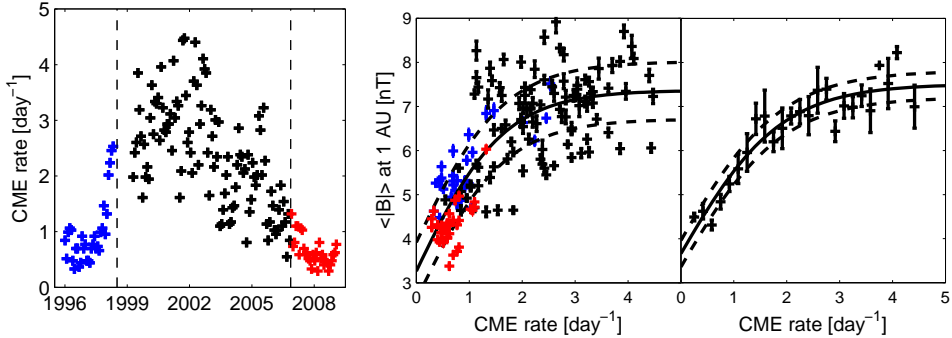}}
\caption{Updated plots of Carrington-Rotation-averaged data from \citet{owe08}. Red (blue) points are from current (previous) solar minimum. Left: CME rate as function of time. Center, right: CME rate versus magnetic field strength at 1 AU. Solid curves are quadratic fits, with dashed curves at 95\% confidence intervals. Values in right panel are binned by CME rate.}
\label{fig:1}
\end{figure}

If the rise and fall of heliospheric field strength is governed solely by the injection of CME loops and the subsequent opening of those loops on time scales of 40-50 days, as estimated by \citet{owe06}, then the rate of injection of CMEs should be proportional to magnetic field strength. The center panel of Figure \ref{fig:1} shows this to be true in an update of the scatter plot of Carrington-Rotation averages of these parameters presented by \citet{owe08}, where the most recent values are from the National Space Science Data Center's OMNI data set and STEREO data at {\tt http://cor1.gsfc.nasa.gov/catalog/}. Although the correlation is fully expected, given the known solar cycle variation of the two parameters, there are two aspects of note. First, the relationship between CME rate and field strength is nonlinear (fit with a quadratic curve), a point relevant to model development addressed in the next section.  Second, for zero CME rate, the field strength has a finite value.

The finite value of field strength for zero CME rate implies that the field has a ``floor'' value \citep[cf.][]{sva07}. In the context of the \citet{owe06} model, this is the value it would reach during solar minimum if all the CME loops introduced during that cycle had opened by interchange reconnection and all that remained was the conserved open flux. According to the smoothed plot in the right panel of Figure 1, the floor value is $\sim$3.7 nT. This is lower than the $4.0\pm 0.3$ nT found by \citet{owe08} owing to the addition of the most recent data points with lower values, but it is still considerably above zero. It seems unlikely that the curve in Figure \ref{fig:1} could steepen enough to pass through the origin. Thus, the unprecedented low values of field strength during the current solar minimum period do not discount the possibility that interchange reconnection of CME loops is the means by which flux is reduced in the heliosphere (option 1, above) and that open flux is conserved.

\section{Discussion}
\begin{figure}
\centerline{\includegraphics[width=0.7\textwidth,clip=]{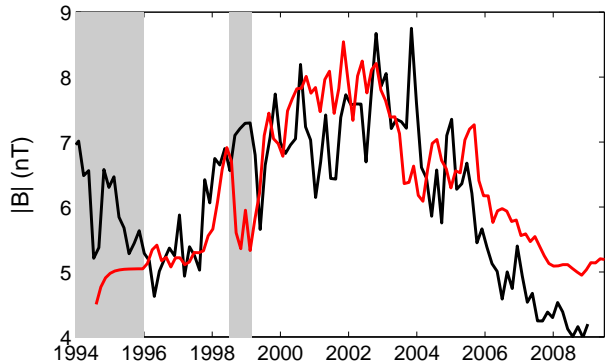}}
\caption{Modeled (red) and observed (black) magnetic field strength at 1 AU as function of time. Shaded intervals indicate gaps in CME rates used in model calculation (after Owens and Crooker, 2006).}
\label{fig:3}
\end{figure}

The \citet{owe06} model implies that the reason the heliospheric field strength did not come close to the deduced floor value in Figure 1 during previous solar minima is that CMEs continued to feed loops into the heliosphere, albeit at a reduced rate, throughout the minima periods. A look at the CME rate as a function of time in the left panel of Figure \ref{fig:1} confirms that CMEs occurred throughout the two minima for which we have CME data. On the other hand, against expectations based upon the \citet{owe06} model, the rate during the current minimum is only marginally lower and more extended than in the previous minimum.  This contrasts with the much more pronounced dip in field strength during the current minimum compared to the previous one, as shown by the black trace in Figure \ref{fig:3}. The contrast is also apparent in the center panel of Figure \ref{fig:1}, where the red points from the current minimum sit at a lower level of field strength than the blue points from the previous minimum. The impact of these contrasting levels at solar minimum on the model is that the model overestimates the field strength during the current minimum, as indicated by the red trace in Figure \ref{fig:3}. The overestimate begins as early as late 2005.

The overestimate of magnetic flux from CMEs by the \citet{owe06} model for the current solar minimum may be owing to its assumption of constant magnetic flux per CME. Recent observations indicate a secular change in CME properties which may reflect a decrease in flux content. Compilers of the automated ``CACTUS'' CME catalog \citep{rob09} noted that CMEs are slower and coming from weaker field regions during the current minimum (E. Robbrecht, private communication, 2009), and A. Vourlidas, (2009 SHINE Workshop) reported a significant drop in mass density.  It is conceivable that CMEs with less mass and speed also have weaker magnetic fields.

Aside from the issue of differences between the current and previous minima, the \citet{owe06} model needs to accommodate the quadratic relationship between magnetic field strength and CME rate shown in Figure \ref{fig:1}. The model relationship is linear owing to the combined assumptions of constant magnetic flux per CME, discussed above, and a constant rate for the interchange reconnection that gradually opens the CME loops. Figure \ref{fig:1} implies that during periods of high CME rate, near solar maximum, each CME increases the field strength by a lesser amount than during periods of low CME rate, near solar minimum. A possible reason for this pattern is that the interchange reconnection which opens the loops occurs at a faster rate during solar maximum, when the solar magnetic field configuration is complicated and evolving rapidly compared to minimum. Changing the model rate of interchange reconnection from constant to a function of solar cycle should bring consistency with Figure \ref{fig:1}.

Finally, we add that the assumption of a constant rate of interchange reconnection may also be a factor in the overestimate of flux from CMEs by the \citet{owe06} model for the current solar minimum, in addition to the assumption of constant flux per CME. The configuration of the solar magnetic field during the declining phase of the solar cycle and into the current minimum was more complicated than during the previous cycle, with higher-order fields dominating the pattern. This configuration may be more conducive to a higher rate of interchange reconnection and, hence, a faster rate of flux loss, compared to the dipole-dominated pattern of the previous minimum.

\section{Conclusions}
The unprecedented low magnetic field strength in the heliosphere during the current solar minimum does not eliminate the possibility that open flux is conserved. Nor does it render invalid the hypothesis that the heliospheric field strength depends solely upon the rate of injection of loops from CMEs and the rate those loops open by interchange reconnection. According to this view, the field floor reflecting the background open flux is never reached during solar minima owing to a residual influx of CME loops. As modeled by \citet{owe06}, however, the extended lower CME rate during the current minimum cannot fully account for the depressed field strength. Assumptions of constant flux per CME and/or rate of interchange reconnection must be relaxed, as seems reasonable owing to apparent changes in CME properties during the current minimum.

\acknowledgements 
The authors thank O. C. St. Cyr for advance values of CME rates.  This research was supported in part by the US National Science Foundation under grant ATM-0553397 and by NASA under grant NNG06GC18G.

\end{document}